# Method for Specifying Location Data Requirements for Intralogistics Applications

Jakob Schyga, Markus Knitt, Johannes Hinckeldeyn, Jochen Kreutzfeldt

*Institute of Technical Logistics, Hamburg University of Technology, Hamburg Germany*

**Abstract**

Various applications leverage location data to increase transparency, efficiency, and safety in intralogistics. There are several properties of location data, such as the data's degrees of freedom, system latency, update rate, or accuracy. To select a suitable indoor localization system, corresponding data requirements must be derived by analyzing the considered application. To date, the dependencies of the system performance and location data requirements have not been satisfactorily described in the literature. Thus, no method exists to adequately derive location data requirements. For intralogistics, such a method is of particular relevance due to the high-cost sensitivity and heterogeneity of partially safety-relevant indoor localization applications. To fill this gap, a method for selecting and quantifying location data requirements for the application in intralogistics is presented in this work, creating substantial added value for warehouse managers and system integrators. The method is based on a spatial model that is built on the premise that location data is used to determine the presence or absence of an entity in a multidimensional interest space. The usage of the method is demonstrated in an exemplary case study for the application of 'Automated Pallet Booking'.

**Keywords**

Indoor localization; Intralogistics; Data requirements; Spatial model; Method

## 1. Introduction

Industry 4.0, Smart Logistics, and Industrial Internet of Things are terms used to describe the ongoing shift in production and logistics driven by modern information and communication technologies. One of the key technologies in intralogistics is indoor localization [1]. Indoor Localization Systems (ILS) generate location data of entities, such as goods, assets, personnel, or vehicles that can be leveraged by a wide range of applications to increase efficiency, safety, transparency, and flexibility in intralogistics. The much-discussed Real-Time Locating Systems (RTLS) are a subcategory of ILS for real-time remote locating. 'Automated Pallet Booking' [2], 'Risk Assessment' [3], and 'Location-Dependent Order Allocation' [4] are just a few examples of the myriad applications of ILS presented in the literature.

For warehouse managers or system integrators to select a suitable ILS for an application, the requirements must be met by the system's performance. Mautz [5] lists the most important requirement parameters for ILS, such as accuracy, integrity, or market maturity. Measurable properties related to the location data itself are considered location data requirements, for which corresponding performance metrics can be determined by experimental evaluation. The results of such experiments are often published in benchmarking studies [6,7] and summarized in literature surveys [5,8]. But how are the data requirements for an application to be determined? As pointed out by Hohenstein and Günthner [9], data requirements depend on the specific implementation and environment of an application. Thus, providing universally applicable figures is not possible. Instead, methodical approaches can be applied to specify data requirements. High-level procedures for the determination of user requirements were proposed by Mautz [5] as well as Gladysz and Santarek [10]. However, when it comes to explicitly selecting and quantifying data requirements, they are not applicable.

In this work, a method is proposed to select and quantify data requirements for intralogistics applications. The remaining work is structured as follows. In Section 2 an overview of related work is presented to provide a basis for the method presented in Section 3. Subsequently, the usage of the method is demonstrated for the application of 'Automated Pallet Booking' in Section 4. Finally, the results are discussed and conclusions are made in Section 5.

## 2. Related work

This section provides an overview of the most relevant work regarding the specification of localization data requirements and concludes by pointing out the identified gaps. Individual indoor localization technologies are deliberately not presented, since data requirements must be considered independently of the functionality.

The most holistic examination of localization requirements is provided by Mautz [5], as part of an extensive literature survey on ILS. Mautz provides a list of 16 well-defined user requirement parameters, divided into the categories 'positioning', 'human-machine interface', 'security and privacy', and 'costs'. Requirement parameters of the 'positioning' category are 'accuracy', 'coverage', 'integrity', 'availability', 'continuity', 'update rate', 'system latency', and 'data output'. In addition, a generic procedure for capturing user requirements is provided. Finally, requirements for selected application domains are exemplarily derived, such as for 'Underground Construction' and 'Ambient Assisted Living'. However, the work lacks to provide information on how the data requirements are ultimately quantified.

Gladysz and Santarek [10] present a procedure for selecting suitable ILS for an application, whereby the initial step deals with the definition of requirements by describing the business case and determining the limits of acceptable requirement parameters. The authors list 'costs/benefits', 'accuracy', and 'reliability' as common requirements, but do not limit the possible parameters to be considered. In addition, a case study of a forklift truck control and diagnostic tool in a cold chain warehouse is presented. The minimum requirement for horizontal position accuracy is specified as 0.5m. Similar to the procedure provided by Mautz [5], the procedure presented by Gladysz and Santarek remains at a high level, without further explanation on how values are quantified.

Hohenstein and Günthner [9] present a survey to examine the suitability of 25 ILS for localizing forklift trucks. The considered evaluation criteria are 'localization accuracy', 'outdoor capability', 'flexibility', and 'scalability'. In addition, 'real-time capability' and 'integration effort' are mentioned as relevant parameters but were not further considered due to a lack of data availability. The localization accuracy is defined as the $95^{th}$ percentile of the horizontal position error. The authors identify the size of the object that must be (indirectly) localized as the relevant criterion to quantify the requirement for localization accuracy. For example, in the course of an automatic booking process, a pallet to be localized must be assigned to a storage location with a known position. The horizontal position accuracy requirement concerning the center of the pallet is accordingly determined by half the width of the storage location. For comparison, the authors give a rough range estimation for the required localization accuracy of five common areas/objects in intralogistics, such as 'storage area', 'storage aisle', and 'storage location'. A generalized method is not provided.

Although the performance requirements of ILS are discussed in several publications, depth is lacking when it comes to quantifying the data requirements. Dependencies of system performance and location data requirements are barely discussed. Hohenstein and Günthner [9] present an interesting approach to determine the requirements for the horizontal position error for the localization of forklift trucks in intralogistics based on the dimensions of the object or area of interest. To create significant added value for warehouse managers and system integrators, this approach must be generalized and further developed.

## 3. Method for deriving location data requirements

In this section, a method is presented for systematically deriving location data requirements. The method is based on the concept of localization functions and comprises the definition of data requirement parameters (Section 3.1), a generic model to describe the spatial dependencies of the data requirement parameters and localization functions (Section 3.2), and a procedure to support the systematical derivation of the defined parameters (Section 3.3).

### 3.1 Localization functions and data requirement parameters

Location data requirements are a subcategory of system requirements that deal with the properties of location data. Location data serve an application to enable localization functions that describe which entity (or entity class) is within (or outside of) a given multidimensional space. This space is considered **Interest Space** and the entity is considered **Entity to be Localized**. If the *Interest Space* is entered or exited, an event is created and transmitted to an application. The application then processes the information to create value for the end-user. Figure 1 illustrates this process for an application with multiple localization functions, ILS, and *Entities to be Localized*.

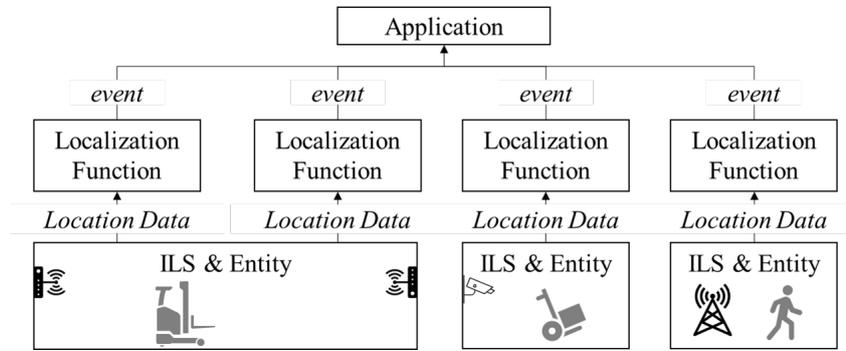

Figure 1: Different localization functions creating events for an application based on the location of different entities. Location data can be provided by one or more ILS and entities to one or more localization functions.

Location data ($\vec{L}$) are a set of discrete values that convey information about an entity's location over time. The term location ($\vec{L_t}$) denotes a position and orientation in space at a certain point in time. Considering a rigid body in three-dimensional space, $\vec{L_t}$ is given by $(x_t, y_t, z_t, \alpha_t, \beta_t, \gamma_t)$, with the position $(x_t, y_t, z_t)$ and orientation $(\alpha_t, \beta_t, \gamma_t)$, whereby the z-axis is corresponding to the vertical direction. Hence, location data consist of up to six **Degrees of Freedom** (DoF). ILS often provide additional data, such as the velocity, acceleration, or the predicted position of an entity. This information can be crucial, for example, in controlling robots. If no additional data is provided, location data may be used for their calculation. When examining location data requirements for the computation of such data, different complex effects must be taken into account [11]. Therefore, the conclusions drawn in this work only apply to localization functions as explained above.

Location data can be differentiated according to the **Localization Type** into *absolute* and *relative localization* [5]. The term absolute localization refers to location estimation in a global frame of reference as defined by landmarks or anchor nodes. In contrast, relative locations are expressed in a local coordinate frame. Figure 2 shows the top view of a logistics scenario with a forklift truck ($O_{forklift}$) and a pallet ($O_{pallet}$) within an area of a warehouse ($O_{area}$). Here, absolute localization refers to the forklift's or the pallet's location in the global reference frame $O_{area}$. Relative localization refers to the forklift's location with respect to the pallet's location or vice versa.

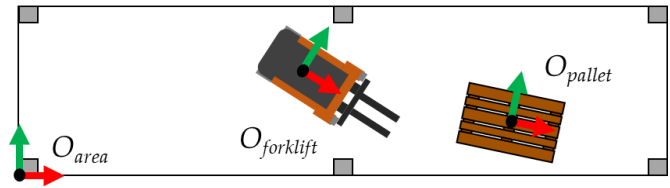

Figure 2: Absolute and relative localization

The location estimate is subject to various error components and thus differs from an entity's true location. The closeness of agreement between the location estimate and the true location is described by the term **Localization Accuracy** [12]. The associated absolute localization error vector is given as the elementwise distance between the location estimate and the true location. Localization errors can be expressed by each element or any combination, such as the horizontal position error.

A parameter closely related to localization accuracy is **Localization Repeatability**. Localization repeatability indicates the closeness of agreement between location estimates at the same true location [12]. Figure 3 visualizes the distinction between accuracy and repeatability. Repeatability can be high even when accuracy is low. In intralogistics applications, localization repeatability is relevant, if the *Interest Space* is specified with respect to location estimates from the same ILS. This is the case, for example, when a robot's navigation is based on a map recorded by the same system [13].

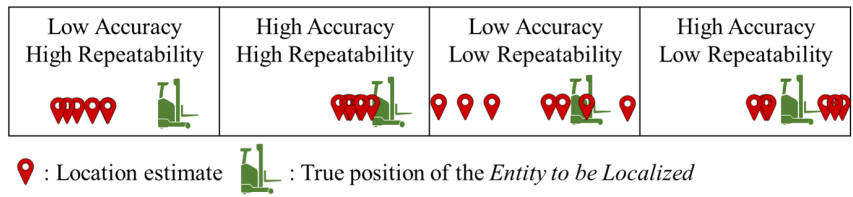

Figure 3: Difference between localization accuracy and repeatability

Relevant metrics for localization accuracy and repeatability are the percentiles of the localization error distribution, which indicate a value below which a certain percentage of location estimates fall. For example, for the horizontal position error, a 95[th] percentile of one meter indicates, that 95 % of location estimates from a set of location data map to a value below this number. Percentiles are essential from an end-user perspective as they indicate the **Confidence** with which a localization accuracy threshold will be met to reliably enable a localization function. The 95[th] percentile became established to define a comparable value for localization accuracy. However, depending on the required confidence of the location estimate, other percentiles can be considered.

Typically, ILS provide **Location Updates** periodically with a constant update rate ($U$). The time gap between two consecutive location measurements is then given by $t_g = 1/U$ (Figure 4). For moving entities, the update rate is relevant since the last location update could indicate the entity is within an *Interest Space* even though the space has already been exited, or vice versa. Besides periodic updates, location updates can be provided *upon request* by an outside trigger or *upon event* triggered by a sensor of the ILS [5].

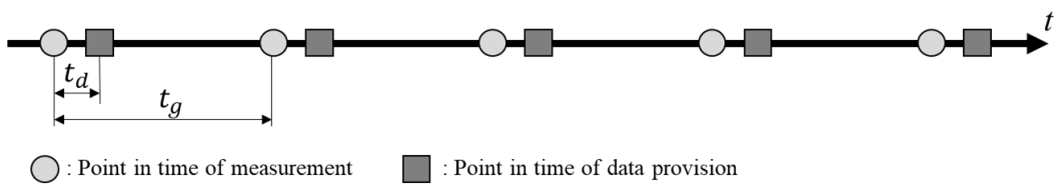

Figure 4: Time gap and system latency

Finally, the **System Latency** describes the time delay ($t_d$) from the actual measurement to the provision of the location data, whereby the timestamp associated with the location estimate is usually set to the measurement point in time. System latency is relevant when location data must be processed near real-time, which is often the case in safety-sensitive applications such as collision avoidance [5].

### 3.2 Spatial model of data requirements

The DoF, localization type, localization accuracy, localization repeatability, location update type, and system latency can be relevant data property parameters to determine the presence of an entity within an *Interest Space*. In this subsection, the dependencies of the data requirement parameters are described in a generic model under consideration of the spatial requirements of localization functions. Finally, the dependencies are expressed in a mathematical equation.

A location update provided by an ILS relates to a specific coordinate frame of the entity to be localized that is usually given by the location of a localization device. This is referred to as the **Entity's Localization Frame** ($O_L$). The coordinate frame that is to be determined as being within an *Interest Space* is referred to as the **Entity's Interest Frame** ($O_I$). Figure 5 illustrates the top view of an *Entity's Localization Frame* and the *Entity's Interest Frame* for a forklift truck. The location estimate of the ILS refers to the rear of the truck, whereby the center between the fork ends of the truck is the relevant location for the localization function. To compute the location of the *Entity's Interest Frame*, the coordinate transformation ($T_{L,I}$) must be applied to the location data provided by the ILS.

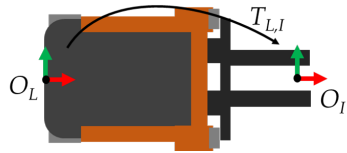

Figure 5: Top view of transformation between the *Entity's Localization Frame* and the *Entity's Interest Frame*

If the localization error at the *Entity's Interest Frame* is too high, the presence or absence of the entity within the *Interest Space* cannot be reliably determined. The closer the *Entity's Interest Frame* is to the boundary of the *Interest Space*, the smaller the localization error must be. The multidimensional space in which the *Entity's Interest Frame* is allowed to move without falsely triggering an event is referred to as **Motion Space**. The closest distance for each of the considered DoF is referred to as **Requirement Margin**. Optionally, a **Safety Margin** can be added. Figure 6 (left) illustrates these spaces for 2-DoF ($x_t, y_t$).

The *Requirement Margin* specifies the maximum localization error that a location estimate at the *Entity's Interest Frame* can have while ensuring the correct determination of presence inside or outside an *Interest Space*. The **Uncertainty Space**, on the other hand, is introduced to describe the localization error at the *Entity's Interest Frame* as a consequence of the system's performance. It refers to the same dimension as the location estimate and comprises the three components shown in Figure 6 (right).

Depending on the relevant reference coordinates for the localization function, **Static Uncertainty** is either determined by the localization accuracy or by the localization repeatability of an ILS, transformed to the

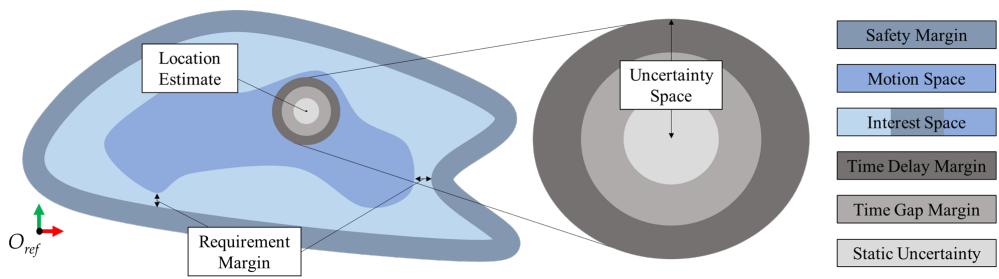

Figure 6: Planar view of spatial components for the *Requirement Margin* and the *Uncertainty Space*

*Entity's Interest Frame.* Rotational error components in the *Entity's Localization Frame* lead to additional translational error components for the location estimation in the *Entity's Interest Frame.* The *Static Uncertainty* ($\vec{P}_{C,Acc/Rep,I}$) at the *Entity's Interest Frame* is provided as the same percentile as the respective accuracy or repeatability of the ILS, which is chosen according to the required confidence level ($C$). This uncertainty component is referred to as static because it is considered to be independent of the entity's motion. This assumption applies to a good approximation at low velocities, depending on the applied localization technology. Time-related uncertainty components are additionally considered for dynamic scenarios. The **Time Gap Margin** refers to the uncertainty component that is resulting from the *Entity's Interest Frame* changing its location since the last location update ($t_g$). The *Time Gap Margin* is not relevant for location updates *upon request* or *upon event*, in which case the data is requested when required. Finally, the **Time Delay Margin** refers to the location uncertainty as a result of the system latency $t_d$. During this time delay, the entity's location changes according to its velocity components. Thus, the *Time Delay Margin* must be considered when the real-time location of the entity is relevant.

The dependencies of the presented model can be expressed in a mathematical equation. The *Requirement Margin* is denoted as a vector $\vec{R}$, with each element relating to a required DoF. The *Interest Space* is denoted as $I$, the *Motion Space* as $M$, and the *Safety Margin* as $S$. The *Uncertainty Space* $\vec{U}$ is a vector of the same dimension as $\vec{R}$ and is given as the sum of the *Static Uncertainty* $\vec{U_s}$, the *Time Gap Margin* $\vec{TG}$, and the *Time Delay Margin* $\vec{TD}$. From the condition that the data requirements must be met by the system performance, it follows

$$\vec{R}(I, M, S) \geq \vec{U_s} + \vec{TG} + \vec{TD}. \tag{1}$$

An upper bound for the time delay and the *Time Gap Margin* results from considering the maximum velocity ($\vec{v}_{max}$) of the *Entity's Interest Frame* multiplied by the maximum time gap or latency. With the accuracy or repeatability percentile for the chosen confidence, it follows

$$\vec{R}(I, M, S) \geq \vec{P}_{C,Acc/Rep,I} + \vec{v}_{max} * t_g + \vec{v}_{max} * t_d. \tag{2}$$

Equation 2 thus leads to a conservative estimate of the requirements. The left side of the equation can be estimated downward and the right side upward, so that the condition of a higher *Requirement Margin* than the *Uncertainty Space* remains satisfied. To quantify location data requirements, the components of this generic equation must be specified in terms of a particular localization function.

### 3.3 Procedure for deriving location data requirements

The presented spatial model and its associated equation form the basis for deriving location data requirements. In this subsection, a procedure with four main steps is presented that is used to specify the localization function (A), determine the *Requirement Margin* (B), estimate the *Uncertainty Space* (C), and finally calculate the data requirements (D). In the following, the steps are briefly explained in the context of intralogistics. The procedure is illustrated in (Figure 7).

**Specify Localization Function (A):** The basis for deriving data requirements is formed by specifying the localization function under consideration. For this purpose, it must be clarified which entity or entity class is to be localized (1) inside or outside which *Interest Space* (2). For example, should a person be localized in front of a shelf or a forklift in an aisle? The localization type (3) can already be deduced from the answer to the question of whether the relevant location data of the *Interest Space* should be specified with respect to the entity's location or a global reference frame. Furthermore, the relevant DoF of the *Interest Space* must be specified (4). Localization functions for intralogistics applications rarely require 6-DoF. Usually, roll and pitch can be neglected, resulting in location data with a maximum of 4-DoF $(x, y, z, \gamma)$. Often, the number of DoF can be further reduced. For example, for ground-based vehicles, considering 3-DoF $(x, y, \gamma)$ and for personnel, the position in the horizontal plane, i.e., the 2-DoF location $(x, y)$, is usually sufficient.

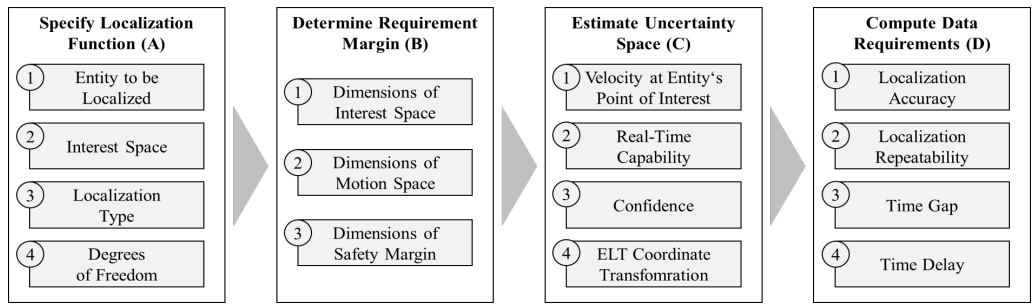

Figure 7: Procedure for deriving location data requirements

**Determine Requirements Margin (B):** In the second step, the *Requirement Margin* is determined. This is done by considering the dimensions of the *Interest Space* (1), the *Motion Space* (2), and the *Safety Margin* (3) for the required DoF. The spatial structure of warehouses and production halls can be well described by rectangular areas or cuboid spaces. For example, an aisle or a storage area in a warehouse can be modeled by rectangles, while a typical storage compartment can be modeled by a cuboid. These abstractions and the resulting symmetries can prove useful in specifying the *Requirement Margin*. For example, if the presence of a forklift in a warehouse aisle is to be determined (localization function), the *Interest Space* is given by the boundaries to the adjacent aisles, including racks. However, the *Motion Space* is limited by the free aisle area and the forklift's dimensions. If the *Entity's Interest Frame* is in the center of the forklift, the boundary of the *Motion Space* is given by the free aisle area (without racks) minus half the width of the forklift.

**Estimate Uncertainty Space (C):** Next, the components of the *Uncertainty Space* are specified, i.e. the right-hand side of Equation 1. If the estimation of the time-related uncertainty components based on the maximum velocity is considered reasonable with respect to the given localization function, Equation 2 can be applied. (1) The maximum velocity ($\vec{v}_{max}$) at the *Entity's Interest Frame* can be estimated considering the application processes. (2) The *Time Delay Margin* ($\overrightarrow{TD}$) must be considered, if the location data must be provided in near real-time. (3) The confidence level ($C$) can be chosen using standard percentiles or according to the *Six Sigma*-method, which has become an established quality management tool in the industry. (4) The coordinate transformation ($T_{L,I}$) between the *Entity's Localization Frame* and the *Entity's Interest Frame* can either be measured or estimated. If the transformation does not contain translational and rotational components, it does not influence the *Static Uncertainty* ($\overrightarrow{U_s}$) and can therefore be neglected.

**Compute Data Requirements (D):** Finally, the data requirement parameters can be calculated. The *Uncertainty Space* can depend on multiple data requirement parameters, such as (1) localization accuracy, (2) localization repeatability, (2) time gap, and (3) time delay. If one uncertainty component is lower, another can be higher. There can be an infinite number of value combinations that satisfy the equation. The specified equation can therefore either be used to calculate different combinations of values or to prove the suitability of an ILS with given data requirement parameters.

4. **Case study**

The proposed method is applied to the exemplary application 'Automated Pallet Booking', which automatically reports the storage or retrieval of pallets in a storage compartment or area to the warehouse management system. This case study serves to demonstrate the feasibility and the usage of the method.

The case study focuses only on the localization function that associates a pallet to a storage compartment of a pallet rack. Thus, the *Entity to be Localized* is a pallet (A, 1) and the *Interest Space* is a storage compartment in a pallet rack (A, 2). Regarding the localization type (A, 3), the location of the storage compartment is known in a global coordinate system. Consequently, the absolute localization of the pallet is required. To determine which compartment a pallet is in, the horizontal and vertical positions are of relevance. Hence, 3-DoF $(x, y, z)$ are considered (A, 4).

Next, the *Requirement Margin* must be determined. Figure 8 shows the front view of an exemplary pallet rack in which pallets are stored lengthwise. To ensure that a pallet can be assigned to the correct compartment, its presence must be distinguished from the adjacent storage compartment. The *Interest Space* is therefore given by a cuboid with the width, depth, and height of the storage compartment (B, 1). The *Motion Space* in $x$, relative to the center of the pallet, results from the space between two adjacent pallets (B, 2). Since the pallet is placed on the crossbar, the $z$-component of the *Motion Space* is zero. Therefore, in Figure 8, the *Motion Space* is visualized as a line in the $xz$-plane. The $x$ and $z$-components of the *Requirement Margin* are marked as $X_R$ and $Z_R$. Additionally, a *Safety Margin* is considered (B, 3). Analogous considerations can be applied to the $y$-component. The calculation of the individual values will not be discussed further, since the focus is on the method itself. For the following derivation of requirements parameters, a *Requirement Margin* of (0.3m, 0.5m, 0.15m) is assumed.

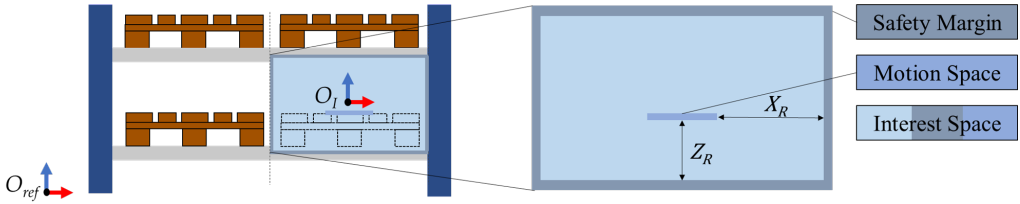

Figure 8: *Interest Space*, *Motion Space* and *Safety Margin* for the application 'automated pallet booking'

The vector for the maximum velocity is assumed to be (0.1 m/s, 0.7 m/s, 0.1 m/s), with substantially higher maximum velocity in the storage direction (C, 1). No real-time capability is required for the localization function (C, 2). Incorrect entries in the warehouse management system lead to various further mistakes and must be avoided. Therefore, a high confidence level (C, 3) of 4σ is exemplarily chosen, which refers to 0.62% false location estimates. The localization device is assumed to be attached at the center of the pallet, which is also the *Entity's Interest Frame*. Thus, the coordinate transformation from the *Entity's Localization Frame* into the *Interest Frame* can be neglected (C, 4). By inserting the values into Equation 2, it follows.

$$\begin{pmatrix} 0.3 \\ 0.5 \\ 0.15 \end{pmatrix} \geq \vec{P}_{4\sigma,Acc/Rep,I} + \begin{pmatrix} 0.1 \\ 0.7 \\ 0.1 \end{pmatrix} \frac{m}{s} * t_g. \quad (3)$$

Since the location of the *Interest Space* is provided in absolute coordinates, localization accuracy is the relevant criterion for the *Static Uncertainty* (D, 1). The localization repeatability can thus be neglected (D, 2). Table 1 shows the resulting 4σ-percentiles of the absolute accuracy for selected values of $t_g$ (D, 3). For an ILS transmitting periodic location updates at an update rate of 2 Hz, corresponding to a time gap of 0.5s, the minimum localization accuracy would therefore be given by $\vec{P}_{4\sigma,Acc}$ equal to (0.25m, 0.15m, 0.12m). Based on these data requirements, an ILS suitable for the considered localization function can be selected.

Table 1: Location data requirements for different time gaps and 4σ-percentiles of the absolute accuracy components

| $t_g$ / [s] | 0.1 | 0.2 | 0.3 | 0.4 | 0.5 | 0.6 |
|---|---|---|---|---|---|---|
| $P_{4\sigma,Acc,x}$ / [m] | 0.29 | 0.28 | 0.27 | 0.26 | 0.25 | 0.24 |
| $P_{4\sigma,Acc,y}$ / [m] | 0.43 | 0.36 | 0.29 | 0.22 | 0.15 | 0.08 |
| $P_{4\sigma,Acc,z}$ / [m] | 0.15 | 0.14 | 0.13 | 0.13 | 0.12 | 0.12 |

## 5. Discussion and conclusions

The potential of leveraging location data in warehouse and production environments is immense as a wide range of applications can be implemented or supported by location data. For warehouse and system integrators, selecting an appropriate ILS is key to implementing a reliable and cost-effective application. Although requirements for ILS have been studied in the literature, the dependencies between the data requirement parameters and spatial requirements have not been adequately described. Thus, a method for the systematic selection and quantification of data requirements does not yet exist. This work fills this research gap by providing a method for systematically deriving location data requirements for intralogistics applications. The method thus supplements the high-level methods for deriving user requirements for ILS as presented by Mautz [5] or Gladysz and Santarek [10]. The presented method comprises the following contributions. (1) Selection and definition of location data requirement parameters that are relevant in terms of the concept of localization functions, (2) a generic model describing the spatial dependencies of the application requirements and data requirement parameters, and (3) a procedure to support the systematic quantification of the data requirement parameters based on the presented model. To demonstrate the applicability of the presented method, a case study on the application of 'Automated Pallet Booking' was presented.

Some limitations of the proposed method remain. First, based on the generic equation of the spatial model (Equation 1), different abstractions were proposed to estimate the individual components. In practice, these should be treated with caution on a case-by-case basis. Often, the exact values are unknown and must be conservatively estimated. Second, the method is based on reliably determining the presence or absence of an entity in an *Interest Space*. This corresponds to a semantic discretization of location data. However, some applications require quasi-continuous location data, for example, to derive dynamic properties. Finally, the method focuses on the data requirement parameters relevant to localization functions. The selection of a system requires the consideration of many more requirements, such as 'size', 'integrity', or 'power supply'.

There are numerous benchmarking studies to evaluate the performance of ILS using various performance metrics. To be meaningful from the end user's perspective, the testing procedure and performance metrics should meet the application requirements. Currently, an application-driven framework is being developed that aims at the meaningful testing and evaluation of ILS [14]. Future work will integrate the discussed concepts and the presented method presented into a holistic approach for application-driven testing and evaluation of ILS.

**Acknowledgements**

This work is funded by the Federal Ministry of Education and Research (BMBF, FKZ: 01IS22047). 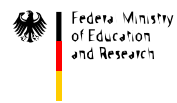

**Biography**

**Jakob Schyga, M.Sc.** (*1992) Research Associate at the Institute for Technical Logistics, Hamburg University of Technology. Jakob Schyga studied mechanical engineering and production at the Hamburg University of Technology from 2012 to 2018. Since 2019, Jakob Schyga has been working on the investigation and application of indoor localization systems in warehouses.

**Markus Knitt, M.Sc.** (*1992) Markus Knitt, M.Sc. (*1997) Research Associate at the Institute for Technical Logistics, Hamburg University of Technology. Markus Knitt studied mechatronics at the Hamburg University of Technology from 2016 to 2022.

**Dr. Johannes Hinckeldeyn** (*1979) Senior Engineer at the Institute for Technical Logistics, Hamburg University of Technology. After completing his doctorate in the UK, Johannes Hinckeldeyn worked as Chief Operating Officer for a manufacturer of laboratory technology for battery research. Johannes Hinckeldeyn studied industrial engineering, production technology, and management in Hamburg and Münster. Current research focuses on the optimization of logistics processes through digital technologies.

**Prof. Dr.-Ing. Jochen Kreutzfeldt** (*1959) Professor and Head of the Institute for Technical Logistics, Hamburg University of Technology. After studying mechanical engineering, Jochen Kreutzfeldt held various management positions at a company group specializing in automotive safety technology. Jochen Kreutzfeldt then took on a professorship for logistics at the Hamburg University of Applied Sciences and became head of the Institute for Product and Production Management. His current research focuses on warehouse and process optimization.